\documentclass[twocolumn,prb,superscriptaddress,longbibliography]{revtex4-2}
\pdfoutput=1
\usepackage{graphicx}
\usepackage{color}
\usepackage{amsmath}
\usepackage{enumitem}
\usepackage{amssymb}
\usepackage[normalem]{ulem}
\usepackage{adjustbox}
\usepackage{bm}
\usepackage{braket}
\usepackage{physics}
\usepackage{dcolumn}
\usepackage{hyperref}
\usepackage{tikz}
\usepackage{gensymb}

\hypersetup{
	colorlinks = true,
	linkcolor = blue,
	citecolor = blue,
	urlcolor  = blue,
}

\DeclareMathAlphabet{\mathdutchcal}{U}{dutchcal}{m}{n}
\SetMathAlphabet{\mathdutchcal}{bold}{U}{dutchcal}{b}{n}
\DeclareMathAlphabet{\mathdutchbcal}{U}{dutchcal}{b}{n}

\begin{document}

\title{Detection of geometric phases in spin waves using nitrogen-vacancy centers}

\author{Tomas T. Osterholt}
\affiliation{%
Institute for Theoretical Physics, Utrecht University, 3584CC Utrecht, The Netherlands\\
}%

\author{Pieter M. Gunnink}
\affiliation{%
Institute of Physics, Johannes Gutenberg-University Mainz, Staudingerweg 7, Mainz 55128, Germany\\
}

\author{Rembert A. Duine}
\affiliation{%
Institute for Theoretical Physics, Utrecht University, 3584CC Utrecht, The Netherlands\\
}
\affiliation{Department of Applied Physics, Eindhoven University of Technology,
P.O. Box 513, 5600 MB Eindhoven, The Netherlands
}%

\date{August 29, 2024}

\begin{abstract}
Due to their robustness, the implementation of geometric phases provides a reliable and controllable way to manipulate the phase of a spin wave, thereby paving the way toward functional magnonics-based data processing devices. Moreover, geometric phases in spin waves are interesting from a fundamental perspective as they contain information about spin wave band structures and play an important role in magnon Hall effects. In this paper, we propose to directly measure geometric phases in spin wave systems using the magnetic field sensing capabilities of nitrogen-vacancy (NV) centers. We demonstrate the general principles of this method on two systems in which spin waves acquire a geometric phase, namely, a wire with a magnetic domain wall and a system with position-dependent anisotropy axes, and explicitly show how this phase can be deduced from the NV center signal. 
\end{abstract}

\maketitle

\section{Introduction} \label{Section Introduction}
For decades significant resources have been devoted to the search for increasingly energy-efficient ways to store and transport data. One active field of research that holds much promise in this regard is the field of spintronics, which gained relevance after the pioneering discovery of giant magnetoresistance by Fert and Gr\"{u}nberg \cite{Fert1988,Grunberg1989}. In recent years research interest in a subfield of spintronics, magnonics \cite{Kruglyak2010,Chumak2015,Pirro2021}, has been growing rapidly. Unlike conventional electronics, information in magnonic devices is transported via spin waves, thus mitigating detrimental effects common in electronic devices such as Joule heating and allowing for more efficient designs of wave-interference-based logic circuits.

For magnonic devices to work properly, however, it is important to have sufficient control over the spin wave phase. Ideally, one would like to be able to manipulate this phase by changing one or more externally controlled parameters of the system under consideration. A natural way to do this would be by implementing geometric phases in the spin waves.  

A general theory for geometric phases in quantum mechanical systems was first given by Berry \cite{Berry1984}, and the concept has proven to be extremely useful in the explanation of phenomena such as the Aharonov-Bohm effect \cite{Berry1984} as well as in theories regarding the anomalous Hall effect \cite{Nagaosa2010,Xiao2010} and topological insulators \cite{Hasan2010,Qi2011}. Geometric phases arise in systems with a Hamiltonian $\hat{\mathcal{H}}(\mathbf{R}(t))$ that depends on some vector $\mathbf{R}(t)$ of parameters which are changing adiabatically in time. These parameters are often quantities that can be controlled externally, such as an applied electric or magnetic field. Now, for such Hamiltonians, the solution to the time-dependent Schr\"{o}dinger equation, $i \hbar \frac{d}{d t} \ket{\Psi(t)} = \hat{\mathcal{H}}(\mathbf{R}(t)) \ket{\Psi(t)} $, is given by \cite{Berry1984}
\begin{equation}
    \ket{\Psi(t)} = \sum_n c_n e^{i\alpha_n(t)}e^{i\varphi_n(t)}\ket{n;t},
\end{equation}
where
\begin{align}
    \alpha_n(t) &= - \frac{1}{\hbar} \int_0^t E_n(t') d t',\\
    \varphi_n(t) &= i \int_{\mathbf{R}(0)}^{\mathbf{R}(t)} \bra{n;t'} \nabla_{\mathbf{R}} \ket{n;t'} \cdot d\mathbf{R}. \label{Equation Geometric Phase QM}
\end{align}
Here $\ket{n;t}$ is an eigenstate of the Hamiltonian $\hat{\mathcal{H}}(\mathbf{R}(t))$ at time $t$ with corresponding energy $E_n(t)$, and the constant coefficients $c_n$ are determined by the initial state $\ket{\Psi(0)}$ and satisfy $\sum_n |c_n|^2 = 1$.  

As is immediately obvious from Eq (\ref{Equation Geometric Phase QM}), the phase $\varphi_n(t)$ depends only on the path traced out by the parameter vector $\mathbf{R}$. When this path forms a closed adiabatic loop in parameter space, $\varphi_n(t)$ is referred to as the \textit{geometric phase} or the Berry phase. Unlike the dynamic phase $\alpha_n(t)$, the geometric phase is sensitive only to the topological structure of the Hamiltonian's parameter space and the geometry of the path $\mathbf{R}(t)$ through this space. Due to this property, geometric phases tend to be robust against random variations in classical control parameters \cite{DeChiara2003}, such as the external magnetic field, which contrasts strongly with the behavior of the dynamic phase.

Geometric phases are not a result specific for quantum mechanics, the generalization of Berry's work to classical systems having been carried out by Hannay \cite{Hannay1985}, and they can therefore also arise in spin wave systems that are described by the semiclassical Landau-Lifshitz-Gilbert equation. This was demonstrated in particular by Dugaev \textit{et al.} \cite{Dugaev2005}, who derived an explicit expression for the geometric phase acquired by spin waves when traveling through textured ferromagnetic structures. Due to the robustness of geometric phases,  it is obvious that they can offer a reliable control over the spin wave phase. However, this is not the only reason why geometric phases in spin waves are of fundamental interest, for they also contain information about the topological properties of the spin wave band structure \cite{Shindou2013,Zhang2013,Mook2014,Owerre2016,Wang2021} and play an important role in magnon Hall effects \cite{Onose2010,Katsura2010,Matsumoto2011,Matsumoto2011_2}. 

Having established the relevance of geometric phases for the field of magnonics, one might now ask how they are to be measured. Although interference is required to observe a Berry phase in quantum mechanical systems, geometric phases in semiclassical systems, and, in particular, geometric phases of spin waves are directly observable quantities and other possibilities for their detection therefore exist. In this paper, we propose to use the magnetic fields generated by the spin waves to directly probe the geometric phase. 

Although these fields are often very weak ($\lesssim 10 \, \mu T$), the technique of nitrogen-vancancy (NV) center magnetometry \cite{Taylor2008,Degen2008,Balasubramanian2008,Maze2008,Rondin2014,Casola2018} offers the required resolution to reliably measure them. As the name suggests, this technique makes use of NV centers, which are point defects in a diamond lattice that consist of a nearest-neighbor pair of a nitrogen atom and a lattice vacancy, to detect magnetic fields. NV centers have in recent years attracted significant attention from the scientific community for other reasons as well, as they hold much potential for the field of quantum computing \cite{Childress2013}. Previous experimental demonstrations of the high magnetic field sensitivity of NV centers include measurements of dispersion relations in yttrium iron garnet (YIG) films \cite{Bertelli2020} and observations of diffusive spin transport in antiferromagnetic insulators \cite{Wang2022}.

In this paper, we theoretically propose that NV center magnetometry can be used to directly probe geometric phases in spin waves without the need for spin wave interference. The remainder of this work is organized as follows. In Sec. \ref{Section NV Center Magnetometry}, we will give a brief summary of the theory behind NV center magnetometry. In Sec. \ref{Section General Theory}, we will then provide a general method to extract the spin wave phase from NV-center-based measurements of spin wave magnetic fields. Finally, in Sec. \ref{Section Examples}, we will show how this technique can be applied to two particular systems, a wire with a magnetic domain wall and a system with position-dependent anisotropy axes, that give rise to a spin wave geometric phase.

\section{NV Center Magnetometry}\label{Section NV Center Magnetometry}
In this section, we briefly introduce NV center magnetometry. Readers familiar with the physics of this technique may wish to skip ahead to the next section.

Although NV centers can exist in different charge states \cite{Mita1996}, only the so-called NV$^-$ state \cite{Doherty2011} can be used efficiently for magnetic field sensing purposes. In this particular state, an additional electron is trapped at the vacancy, which results in the creation of an effective spin-$1$ pair consisting of this trapped electron and one of the vacancy electrons. In the presence of a magnetic field $\mathbf{B}$, the electronic ground state Hamiltonian of the NV$^-$ state is then given by \cite{Alegre2007,Doherty2012,Bertelli2021}
\begin{equation}\label{Equation Ground State Hamiltonian}
    \hat{\mathcal{H}} = D \hat{S}_{z'}^2 + \eta \mathbf{B} \cdot \mathbf{S}.
\end{equation} 
Here $\mathbf{S} = (\hat{S}_{x'},\hat{S}_{y'},\hat{S}_{z'})$ is used to denote the spin-1 operators, with $\hat{\mathbf{z}}'$ being the vector pointing from the vacancy site to the nitrogen atom. Note that we use primed coordinates to indicate the coordinate system with an axis along the NV center, and nonprimed coordinates to denote the laboratory frame. The constants $D \approx 2.87$ GHz and $\eta \approx 28$ GHz/T appearing in the expression correspond to respectively the field-independent and the field-dependent splitting between the three ground state levels.

Given a static magnetic field $\mathbf{B}_0$, one can easily calculate the energies of this Hamiltonian. This process can also be inverted to obtain $\mathbf{B}_0$ from the energy differences between the ground state levels, \footnote{Strictly speaking, you can only obtain the magnitude of the static magnetic field components parallel and perpendicular to the NV axis.} which forms the basis of the technique of NV center magnetometry. These energy differences can be determined experimentally by means of optically detected magnetic resonance measurements, the details of which can be found in Ref. \cite{Bertelli2021}.

Now, the magnetic field generated by a spin wave will not be static. It is still possible, however, to use the NV centers to obtain information about time-varying fields, as we will now discuss. First, one assumes that the total magnetic field $\mathbf{B}$ at the NV center consists of both a harmonically time-varying field part and an applied static field part. That is, we take
\begin{equation}
    \mathbf{B} = (B_R + B_L) \cos(\omega t) \hat{\mathbf{x}}' + (B_R - B_L) \sin(\omega t) \hat{\mathbf{y}}' + B_0 \hat{\mathbf{z}}',
\end{equation}
where the time-varying field with frequency $\omega$ has been decomposed into right-handed ($B_R$) and left-handed ($B_L$) components. Furthermore, we assume that the magnitude of the static field is much larger than that of the time-varying field, $|B_0| \gg \sqrt{B_R^2+B_L^2}$. In the absence of the time-varying field, the stationary states of the ground state Hamiltonian of Eq.~\eqref{Equation Ground State Hamiltonian} are given by $\ket{S m_{z'}}$, with $S=1$ and $m_{z'} = 0, \pm 1$. Abbreviating these levels by $\ket{m_{z'}}$, we note that transitions $\ket{0} \leftrightarrow \ket{\pm 1}$ occur when the magnitude of this time-varying field becomes nonzero; the magnetic-dipole-forbidden transition $\ket{1} \leftrightarrow \ket{-1}$ cannot take place. As shown in Appendix A, the corresponding transition (or Rabi) frequency $\Omega_{R,+}$ ($\Omega_{R,-}$) of the $\ket{0} \leftrightarrow \ket{1}$ ($\ket{0} \leftrightarrow \ket{-1}$) transition depends on both the field frequency $\omega$ and the magnitude of the component $B_R$ ($B_L$). When $\omega$ matches the electronic resonance frequency $\omega_+$ ($\omega_-$), with $\omega_{\pm} = D \pm \eta B_0$ corresponding to the energy difference between the levels $\ket{0}$ and $\ket{1}$ ($\ket{-1}$), the transitions are driven most efficiently and the corresponding on-resonance Rabi frequency $\Omega_{R,+}$ ($\Omega_{R,-}$) is given by \cite{Bertelli2021}:
\begin{align}\label{Equation Rabi Frequency Formula}
    \Omega_{R,+} &= \sqrt{2} \eta |B_R|, \nonumber\\
    \Omega_{R,-} &= \sqrt{2} \eta |B_L|.
\end{align}
A measurement of the Rabi frequencies thus allows one to deduce the magnitudes of the time-varying magnetic field components $B_R$ and $B_L$, which in turn can be used to detect geometric phase effects in spin waves as we show in Section \ref{Section General Theory}. To conclude this section we mention that the Rabi frequencies can be obtained from a simple pulsed control detection scheme, the details of which can be found in Ref. \cite{Bertelli2021}.

\section{General Theory} \label{Section General Theory}
We consider an infinite planar magnetic thin film oriented parallel to the $xy$-plane and centered at $z=0$. We assume that a spin wave with wave vector $\mathbf{k} = k_x \hat{\mathbf{x}} + k_y \hat{\mathbf{y}}$ and frequency $\omega$ travels coherently through the film. Defining $\boldsymbol{\rho} = x \hat{\mathbf{x}}+y \hat{\mathbf{y}}$, we assume that the spin wave accumulates a phase $\varphi(\boldsymbol{\rho})$ such that the magnetization $\mathbf{M}(\mathbf{r},t)$ of the spin wave is given by
\begin{equation}\label{Equation magnetization}
    \mathbf{M}(\mathbf{r},t) 
    =
    \begin{pmatrix}
        m_x(z) \cos[\mathbf{k} \cdot \boldsymbol{\rho}-\omega t + \varphi(\boldsymbol{\rho})]\\
        0\\
        -m_z(z) \sin[\mathbf{k} \cdot \boldsymbol{\rho}-\omega t + \varphi(\boldsymbol{\rho})]
    \end{pmatrix}
    .
\end{equation}
Here $m_x(z)$ and $m_z(z)$ are spin wave amplitudes that can vary as a function of $z$. Note that at this stage we do not make any assumptions as regards the nature of the phase $\varphi(\boldsymbol{\rho})$; it can be a geometric phase, but it need not be. 

We proceed to calculate the spin wave magnetic field just outside the film and show how the phase $\varphi(\boldsymbol{\rho})$ can be extracted from this field by means of NV center magnetometry.

\subsection{Spin Wave Magnetic Field}
Since no free current density is present and the time variation of the electric displacement field is expected to be small, the magnetizing field $\mathbf{H}$ generated by the spin wave can be calculated from the magnetic scalar potential, yielding \cite{Jackson1999}
\begin{equation}\label{Equation Magnetizing Field Formula}
    H_{\mathbf{\beta}} (\mathbf{r},t) = \frac{1}{4 \pi} \partial_{\beta} \int d\mathbf{r}' \frac{\nabla ' \cdot \mathbf{M}(\mathbf{r}',t)}{|\mathbf{r}-\mathbf{r}'|},
\end{equation}
with $\beta= \{x,y,z\}$. Starting from the magnetization, Eq.~\eqref{Equation magnetization}, we find upon changing variables, $\boldsymbol{\rho}_R = \boldsymbol{\rho}' - \boldsymbol{\rho} $, that the exact expression for the magnetizing field of the spin wave is given by
\begin{align}\label{Equation Exact Magnetizing Field}
     H_{\beta} (\mathbf{r},t) &= -\partial_{\beta} \int d \boldsymbol{\rho}_R d z' \biggr(\frac{ \Tilde{k}_x(\boldsymbol{\rho}_R+\boldsymbol{\rho}) m_x(z')+\frac{\partial m_z (z')}{\partial z'} }{4 \pi|\boldsymbol{\rho}_R+(z'-z)\hat{\mathbf{z}}|} \biggr) \nonumber\\ &\times\sin [\mathbf{k} \cdot (\boldsymbol{\rho}_R+\boldsymbol{\rho})-\omega t + \varphi(\boldsymbol{\rho}_R+\boldsymbol{\rho})].
\end{align}
where we have defined the \textit{phase-translated wave vector} $\Tilde{\mathbf{k}}(\boldsymbol{\rho})$ as
\begin{equation}
    \Tilde{\mathbf{k}}(\boldsymbol{\rho}) = \mathbf{k} + \nabla \varphi(\boldsymbol{\rho}). 
\end{equation} 

In general, a simple expression for Eq.~\eqref{Equation Exact Magnetizing Field} cannot be obtained nor can it be expected that there exists a method valid for arbitrary positions $\mathbf{r} = (\boldsymbol{\rho},z)$ to extract information about the local phase $\varphi(\boldsymbol{\rho})$ from the spin wave magnetic field. A solution to both problems is achieved by considering the field $\mathbf{H}(\mathbf{r},t)$ for $\mathbf{r}$ sufficiently close to the film. In this limit effectively all contributions to the magnetic field detected at $\mathbf{r}$ will come from a small region of the film centered around $\boldsymbol{\rho}$, which allows us to use the following approximations inside the integral of Eq.~\eqref{Equation Exact Magnetizing Field}:
\begin{align}
    \varphi(\boldsymbol{\rho}_R+\boldsymbol{\rho}) &\approx \varphi(\boldsymbol{\rho}) + \nabla \varphi (\boldsymbol{\rho}) \cdot \boldsymbol{\rho}_R, \label{Equation First Approximation} \\
    \Tilde{\mathbf{k}}(\boldsymbol{\rho}_R+\boldsymbol{\rho}) &\approx \Tilde{\mathbf{k}}(\boldsymbol{\rho}). \label{Equation Second Approximation}
\end{align}
We note that these approximations are not independent of one another as Eq.~\eqref{Equation Second Approximation} follows directly from Eq.~\eqref{Equation First Approximation}. Defining $\Tilde{k}(\boldsymbol{\rho}) = |\Tilde{\mathbf{k}}(\boldsymbol{\rho})|$ and making use of a Fourier identity \cite{Jackson1999},
\begin{equation}\label{Equation Fourier Identity}
    \frac{1}{|\mathbf{r}-\mathbf{r}'|} = \frac{1}{2 \pi} \int d^2 \mathbf{q} \frac{e^{-q|z-z'|}}{q} e^{i \mathbf{q} \cdot (\boldsymbol{\rho}-\boldsymbol{\rho}')},
\end{equation}
we obtain the following simple expression for the magnetizing field just outside the film:
\begin{equation}\label{Equation Approximate Magnetizing Field}
    \mathbf{H}(\mathbf{r},t) = \frac{M_{\mathbf{k}}(\mathbf{r})}{\Tilde{k}(\boldsymbol{\rho})}
    \begin{pmatrix}
        \Tilde{k}_x(\boldsymbol{\rho}) \cos[\mathbf{k}\cdot\boldsymbol{\rho}-\omega t + \varphi(\boldsymbol{\rho})] \\
        \Tilde{k}_y(\boldsymbol{\rho}) \cos[\mathbf{k}\cdot\boldsymbol{\rho}-\omega t + \varphi(\boldsymbol{\rho})] \\
        - \text{sgn}(z)  \Tilde{k}(\boldsymbol{\rho}) \sin[\mathbf{k}\cdot\boldsymbol{\rho}-\omega t + \varphi(\boldsymbol{\rho})]
    \end{pmatrix}
    ,
\end{equation}
where
\begin{align}
    M_{\mathbf{k}}(\mathbf{r}) &= \Tilde{k}(\boldsymbol{\rho}) \text{sgn}(z) \int d z' \frac{e^{-\Tilde{k}(\boldsymbol{\rho})|z-z'|}}{2}  m_z(z') \nonumber \\
    &-\Tilde{k}_x(\boldsymbol{\rho}) \int d z' \frac{e^{-\Tilde{k}(\boldsymbol{\rho})|z-z'|}}{2}  m_x(z').
\end{align}
To be consistent with our approximations we have neglected all derivatives of $\varphi(\boldsymbol{\rho})$ higher than first order when performing the differentiation with respect to $\beta$. Furthermore, we have used that $\text{sgn}(z-z') = \text{sgn}(z)$ outside the film.

In what follows, we will assume that Eq.~\eqref{Equation Approximate Magnetizing Field} can be used. To check whether in a given situation the approximation scheme leading to Eq.~\eqref{Equation Approximate Magnetizing Field} is applicable, one can always compare the result of Eq.~\eqref{Equation Approximate Magnetizing Field} with a numerical evaluation of the exact result, Eq.~\eqref{Equation Exact Magnetizing Field}. 

Finally, we mention that the spin wave magnetic field $\mathbf{B}(\mathbf{r},t)$ outside the film is related to its magnetizing field via the simple relation $\mathbf{B}(\mathbf{r},t) = \mu_0 \mathbf{H}(\mathbf{r},t)$, with $\mu_0$ the vacuum magnetic permeability.

\subsection{Rabi Frequency}
\begin{figure}
    \centering
    \includegraphics[scale=0.6]{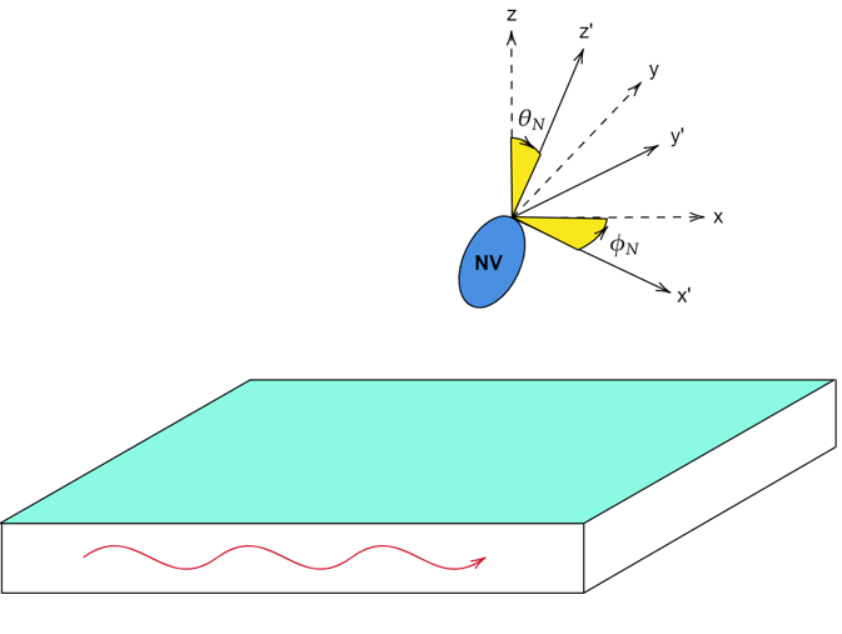}
    \caption{Schematic figure of a single NV center placed above a magnetic film through which a spin wave is traveling. The NV axis ($z'$) coincides with the major axis of the blue ellipse in this figure. Also shown are the laboratory coordinates $\{x,y,z\}$ and the NV center coordinates $\{x',y',z'\}$.} 
    \label{Figure General Setup}
\end{figure}

Having obtained an expression for the spin wave magnetic field, we now proceed to calculate its effect on the NV centers. We consider a family of NV centers with their axis oriented along the unit vector $\hat{\mathbf{z}}'$, where $\hat{\mathbf{z}}'$ is an element of the right-handed orthonormal basis $\{ \hat{\mathbf{x}}', \hat{\mathbf{y}}', \hat{\mathbf{z}}'\}$,
\begin{align}
    \hat{\mathbf{x}}' &= \cos \theta_N \cos \phi_N \hat{\mathbf{x}}+\cos\theta_N \sin\phi_N \hat{\mathbf{y}}-\sin\theta_N \hat{\mathbf{z}}, \nonumber\\
    \hat{\mathbf{y}}' &= -\sin\phi_N \hat{\mathbf{x}}+ \cos\phi_N \hat{\mathbf{y}}, \nonumber \\
    \hat{\mathbf{z}}' &= \sin\theta_N \cos\phi_N \hat{\mathbf{x}}+\sin\theta_N \sin\phi_N \hat{\mathbf{y}}+\cos\theta_N \hat{\mathbf{z}}.
\end{align}
Here $\theta_N \in [0, \pi]$ and $\phi_N \in [0, 2 \pi)$ are angles describing the orientation of the NV axis with respect to the magnetic film basis $\{ \hat{\mathbf{x}}, \hat{\mathbf{y}}, \hat{\mathbf{z}}\}$. Graphically, this coordinate system is shown in Fig. \ref{Figure General Setup}. We note that in this figure only a single NV center is illustrated, although it is in principle straightforward to use an ensemble of NV centers for the measurement \footnote{When dealing with an ensemble, one of course has to deal with the four different possible orientations for the NV centers. However, one can make sure that only one of these orientations contributes to the measured signal by aligning the applied static magnetic field along this particular NV axis. The NV centers oriented along the three other NV axes, which due to the crystal symmetry all make an angle of $71\degree$ with respect to this particular axis, will consequently have a resonance frequency $\Bar{\omega}_{\pm}$ that differs from the resonance frequency $\omega_{\pm}$ of the NV centers oriented along the magnetic field direction, thus allowing one to isolate the contributions coming from this particular orientation.}.

A static uniform magnetic field $\mathbf{B}_0$ is applied parallel to $\hat{\mathbf{z}}'$, which sets the electronic resonance frequencies $\omega_{\pm}$ of the NV centers equal to $D \pm \eta B_0$. Tuning $B_0$ such that either $\omega_+$ or $\omega_-$ coincides with the spin wave frequency $\omega$, the spin wave magnetic field will induce transitions $\ket{0} \leftrightarrow \ket{\pm 1}$ with corresponding Rabi frequencies $\Omega_{R,+}$ and $\Omega_{R,-}$ respectively. Inspired by earlier experiments that were concerned with the detection of the wave vector $\mathbf{k}$ \cite{Bertelli2020}, we require the spin wave magnetic field to interfere with a spatially homogeneous reference field of the same frequency $\omega$. This will result in a spatially-varying profile of the NV center Rabi frequencies which allows one to easily extract the phase $\varphi(\boldsymbol{\rho})$. 

We thus take the reference field to have the following form:
\begin{equation}\label{Equation Reference Field}
    \mathbf{H}_{\rm ref} (\mathbf{r},t) = H^{\rm ref}_{x'} \cos (\omega t) \hat{\mathbf{x}}' + H^{\rm ref}_{y'} \sin (\omega t) \hat{\mathbf{y}}',
\end{equation}
where we take $H^{\rm ref}_{x'}$ and $H^{\rm ref}_{y'}$ to have the same sign; a change in the relative sign will force us to replace the LHS of Eq.~\eqref{Equation General Rabi Frequencies} by $\Omega_{R,\mp}(\mathbf{r})$. The Rabi frequencies resulting from the combined field $\mathbf{H}_I (\mathbf{r},t) = \mathbf{H}_{\rm ref} (\mathbf{r},t) + \mathbf{H} (\mathbf{r},t)$ are then given by:
\begin{align} \label{Equation General Rabi Frequencies}
    \Omega_{R,\pm}(\mathbf{r}) &= \frac{\eta \mu_0}{\sqrt{2}}\biggr| H^{\rm ref}_{x'} \pm H_{y'}^{\rm ref} + \mathcal{A}_1(\mathbf{r}) \cos[\mathbf{k}\cdot \boldsymbol{\rho}+\varphi(\boldsymbol{\rho})] \nonumber\\ &\pm \big(\mathcal{A}_2(\mathbf{r}) \mp \mathcal{A}_3(\mathbf{r}) \big)\sin[\mathbf{k}\cdot \boldsymbol{\rho}+\varphi(\boldsymbol{\rho})]\biggr|,
\end{align}
where we have introduced the following quantities:
\begin{align}
    \mathcal{A}_{1}(\mathbf{r}) &= \frac{M_{\mathbf{k}}(\mathbf{r})}{\Tilde{k}(\boldsymbol{\rho})} \big( \Tilde{k}_x(\boldsymbol{\rho})\cos\phi_N+\Tilde{k}_y(\boldsymbol{\rho}) \sin\phi_N  \big) \cos \theta_N, \nonumber    \\
    \mathcal{A}_{2}(\mathbf{r}) &= \frac{M_{\mathbf{k}}(\mathbf{r})}{\Tilde{k}(\boldsymbol{\rho})} \big( -\Tilde{k}_x(\boldsymbol{\rho})\sin\phi_N+\Tilde{k}_y(\boldsymbol{\rho}) \cos\phi_N  \big),  \nonumber\\
    \mathcal{A}_{3}(\mathbf{r}) &= - \text{sgn}(z) M_{\mathbf{k}}(\mathbf{r}) \sin\theta_N.
\end{align}
A complete derivation of the Rabi frequencies is given in Appendix A. Here we only mention that this expression has been derived under the assumption that the reference field components are much larger than their spin wave field counterparts, $|H^{\rm ref}_{x'}| \gg |H_{x'}|$ and $|H^{\rm ref}_{y'}| \gg |H_{y'}|$. This has the additional benefit that the effect of noise on the NV centers tends to be reduced by larger reference fields \cite{Carmiggelt2023,Cai2012}, thereby improving the sensitivity of the measurement technique. 

As discussed in Sec. \ref{Section NV Center Magnetometry}, the Rabi frequencies of an NV center can be measured directly by means of a pulsed experiment. Placing the NV centers in a parallel plane above or below the film makes the extraction of the phase $\varphi(\boldsymbol{\rho})$ from the detected Rabi frequencies particularly simple, especially in those regions where $k \gg |\nabla \varphi(\boldsymbol{\rho})|$ and hence where $\mathcal{A}_1(\mathbf{r})$, $\mathcal{A}_2(\mathbf{r})$ and $\mathcal{A}_3(\mathbf{r})$ are effectively independent of $\boldsymbol{\rho}$. 

\section{Examples} \label{Section Examples}
Having developed the general theory for the detection of spin wave phases using the technique of NV center magnetometry, we now apply our methodology to two particular systems in which spin waves acquire a geometric phase contribution. The systems that we will consider are a wire with a magnetic domain wall and a magnetic film with position-dependent anisotropy axes. 

Before discussing these examples in detail, we again emphasize that geometric phases arise only when the system's parameter vector $\mathbf{R}$ is taken around a closed adiabatic loop. Within the current context the adiabaticity condition implies that the spatial variations of the system's parameters are required to be small on length scales comparable to the wavelength of the spin wave.

\subsection{Magnetic domain wall} \label{Subsection Magnetic Domain Walls}
\begin{figure}
    \centering
    \includegraphics[scale=0.53]{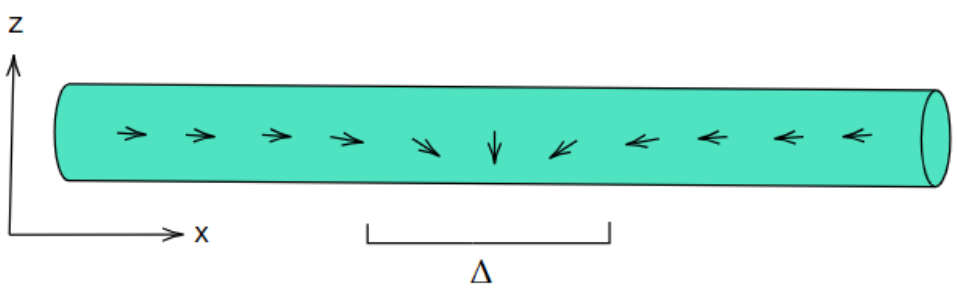}
    \caption{A magnetic wire with a head-to-head domain wall. The direction of the equilibrium magnetization $\hat{\mathbf{m}}_0$ is constrained to the $xz$ plane and illustrated by the arrows. Domain wall widths $\Delta$ with sizes of hundreds of nanometers have been realized in past experiments \cite{Parkin2008}.}
    \label{Figure Magnetic Domain Wall}
\end{figure}

We consider a planar magnetic thin film oriented parallel to the $xy$ plane and extending from $z=-d$ to $z=d$, with $d$ small. The length $L_x \gg d$ of the film in the $x$ direction is taken to be much larger than the corresponding length $L_y \gg d$ in the $y$ direction, and our system is thus effectively described as a magnetic wire. A static head-to-head magnetic domain wall with easy axis along $x$ is present and centered at $x=0$. We assume that the direction $\hat{\mathbf{m}}_0$ of the equilibrium magnetization changes in a smooth manner when crossing this domain wall, pointing toward $+\hat{\mathbf{x}}$ to the left ($x<0$) of the domain wall and pointing toward $-\hat{\mathbf{x}}$ to the right ($x>0$). As is illustrated in Fig. \ref{Figure Magnetic Domain Wall}, we take, for algebraic convenience, the direction of $\hat{\mathbf{m}}_0$ to be constrained to the $xz$ plane. We note, however, that the results derived below can readily be generalized to arbitrary planes.

A more formal description of our system can be given as follows. Defining $\mathbf{m} = \mathbf{M}/M_s$, with $M_s$ the saturation magnetization, we take the energy functional of the system to be given by
\begin{equation}\label{Equation Energy Functional Domain Wall}
    E[\mathbf{m}] = \int d \mathbf{r}' \biggr[ -\frac{J}{2} \mathbf{m} \cdot \nabla^2 \mathbf{m} - \frac{K_x }{2} m_x^2 + \frac{K_y}{2} m_y^2 \biggr].
\end{equation}
Here $J >0$ is the exchange coefficient, and $K_x > 0$ and $K_y \geq 0$ are the anisotropy coefficients in the $x$ and $y$ directions, respectively. The equilibrium magnetization direction $\hat{\mathbf{m}}_0 = (\sin \theta \cos \phi, \sin \theta \sin \phi, \cos \theta)$ is described by a Walker profile in the $xz$ plane \cite{Schryer1974},
\begin{align}
    \phi(\mathbf{r}) &= 0,  \nonumber\\ 
    \theta(\mathbf{r}) &= \theta(x) = \frac{\pi}{2} + 2 \arctan \big( e^{x/\Delta} \big),
\end{align}
with $\Delta=\sqrt{J/K_x}$ the domain wall width.

We seek to study spin waves in the bulk \footnote{With bulk we mean the regions far away from the edges of the film that run parallel to the $x$ and $y$ axes.} that travel straight through this domain wall. It is convenient to introduce spherical coordinate vectors $\hat{\boldsymbol{\theta}}(\mathbf{r}) = (\cos \theta(x), 0,-\sin \theta(x))$ and $\hat{\boldsymbol{\phi}}(\mathbf{r}) = \hat{\mathbf{y}}$ such that we have an orthonormal right-handed basis $\{\hat{\mathbf{m}}_0, \hat{\boldsymbol{\theta}}, \hat{\boldsymbol{\phi}}  \}$ at every position $\mathbf{r}$. We can then write
\begin{align}
    \mathbf{m}(x,t) &\approx \hat{\mathbf{m}}_0(x) + m_{\theta}(x,t) \hat{\boldsymbol{\theta}}(x) + m_{\phi}(x,t) \hat{\boldsymbol{\phi}}(x),
\end{align}
where the magnitudes of the spin wave components $m_{\theta}(x,t)$ and $m_{\phi}(x,t)$ are assumed to be much smaller than $1$.

In the absence of damping, the evolution of $\mathbf{m}$ is governed by the Landau-Lifshitz equation,
\begin{equation}
    \frac{\partial \mathbf{m}}{\partial t} = - \gamma \mathbf{m} \times \mathbf{H}_{\mathrm{eff}},
\end{equation}
with $\gamma$ the gyromagnetic ratio and $\mathbf{H}_{\mathrm{eff}} (\mathbf{r},t)$ the effective field,
\begin{align}
    \mathbf{H}_{\mathrm{eff}} (\mathbf{r},t) &= - \frac{1}{M_s}\frac{\delta E[\mathbf{m}(\mathbf{r},t)]}{\delta \mathbf{m}(\mathbf{r},t) } \nonumber \\&= \frac{J}{M_s} \nabla^2 \mathbf{m} + \frac{K_x}{M_s} m_x \hat{\mathbf{x}} - \frac{K_y}{M_s} m_y \hat{\mathbf{y}}.
\end{align}
Keeping only terms up to first order in $m_{\theta}$ and $m_{\phi}$ in the Landau-Lifshitz equation, we arrive at the following pair of coupled partial differential equations:
\begin{align}
     \frac{M_s}{ \gamma}\frac{\partial m_{\theta}}{\partial t} &=   J \frac{\partial^2 m_{\phi}}{\partial x^2} + K_x \biggr( 2 \cos^2 \theta -1\biggr) m_{\phi}  - K_y m_{\phi}, \nonumber\\
     \frac{M_s}{\gamma}\frac{\partial m_{\phi}}{\partial t} &= -J \frac{\partial^2 m_{\theta}}{\partial x^2} - K_x \biggr( 2 \cos^2 \theta -1 \biggr) m_{\theta}.
\end{align}
In the limit where $K_y$ becomes vanishingly small, \footnote{A nonzero value for $K_y$ is still needed to ensure the stability of domain wall.} these equations can be solved analytically, which yields the following solutions \cite{Yan2011}:
\begin{align}\label{Equation Domain Wall Spin Wave Magnetization}
    m_{\theta}(x,t) &= \Re \biggr\{ C_1 \biggr[ \tanh \biggr( \frac{x}{\Delta}\biggr) - i \, \Delta \, k \biggr] e^{i (k x-\omega t)} \nonumber \\&+ C_2 \biggr[ \tanh \biggr( \frac{x}{\Delta}\biggr) + i \, \Delta \, k \biggr] e^{-i (k x+\omega t)} \biggr \}, \nonumber\\
    m_{\phi}(x,t) &= \Re \biggr\{ i C_1 \biggr[ \tanh \biggr( \frac{x}{\Delta}\biggr) - i \, \Delta \, k \biggr] e^{i (k x-\omega t)} \nonumber \\&+ i C_2 \biggr[ \tanh \biggr( \frac{x}{\Delta}\biggr) + i \, \Delta \, k \biggr] e^{-i (k x+\omega t)} \biggr \},    
\end{align}
where $C_1, C_2 \in \mathbb{C}$ are constants that are related to the spin wave amplitudes. The dispersion relation is given by $ \omega(k) = \gamma (J k^2 + K_x)/M_s$. 

\begin{figure}
    \centering
    \includegraphics[scale=0.56]{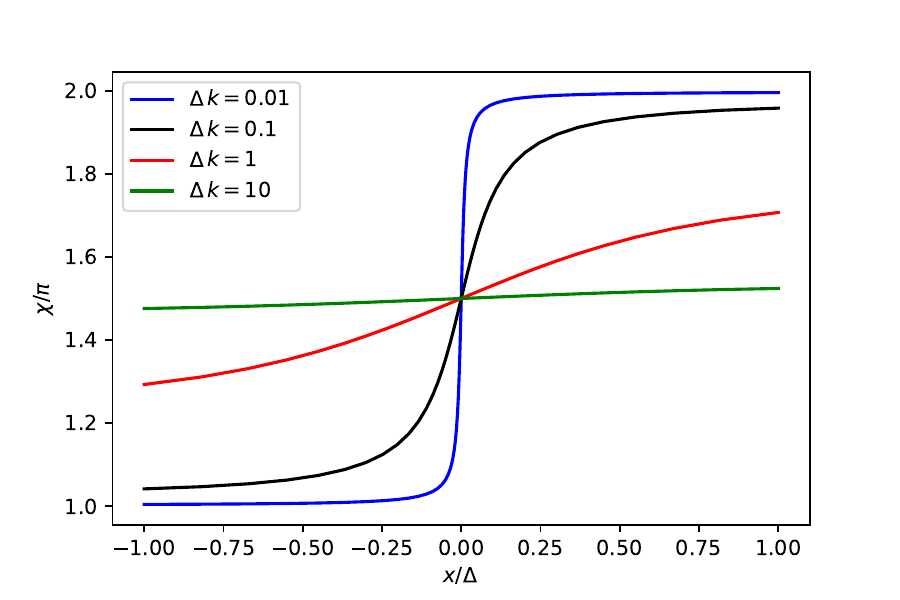}
    \caption{Plot of the phase $\chi(x)$ for different values of $\Delta \, k$. In the adiabatic limit, $k \gg \Delta^{-1}$, $\chi(x)$ becomes position-independent.}
    \label{Figure Chi}
\end{figure}

We will now focus on the case of a spin wave traveling from the left side of the film to the right. That is, we take $k>0$, $C_1 = |m| \neq 0$ and $C_2 = 0$. Expressing the spin wave magnetization in the magnetic film basis and using a trigonometric identity to write it in a more useful form, we find
\begin{align}\label{Equation Final Spin Wave Magnetization Domain Wall}
    \mathbf{M}_{sw}(\mathbf{r},t) =  M(x,z)
    \begin{pmatrix}
        &\cos \theta (x) \cos [ k x - \omega t + \chi(x)] \\
        &-\sin [ k x - \omega t + \chi(x)] \\
        &-\sin \theta (x) \cos [ k x - \omega t + \chi(x)]
    \end{pmatrix}
    .
\end{align}
The phase term $\chi(x)$ is given by
\begin{align}\label{Equation Dynamic Phase Domain Wall}
    \chi(x) &= 
    \begin{cases}
        \pi + \arctan[-\Delta \, k \coth \big(\frac{x}{\Delta}\big)], \hspace{0.45cm} \text{for} \hspace{0.3cm} x \leq 0. \\
        2\pi + \arctan[- \Delta \, k \coth \big(\frac{x}{\Delta}\big)], \hspace{0.3cm} \text{for} \hspace{0.3cm} x > 0.
    \end{cases}
\end{align}
Plots of the phase $\chi (x)$ for different values of the wave vector $k$ are shown in Fig. \ref{Figure Chi}. The function $M(x,z)$ is given by
\begin{align}\label{Equation Magnetization Function Domain Wall}
    \frac{M(x,z)}{M_s|m|} &= \sqrt{\tanh^2\biggr( \frac{x}{\Delta}\biggr)+\Delta^2 k^2} \big[ H(z+d)-H(z-d)\big],
\end{align}
with $H(z)$ the Heaviside step function. Here we have introduced the $z$-dependency of the spin wave magnetization by assuming exchange boundary conditions, from which it follows that the solution homogeneous in $z$ is the lowest-energy state.

\begin{figure}
    \centering
    \includegraphics[scale=0.56]{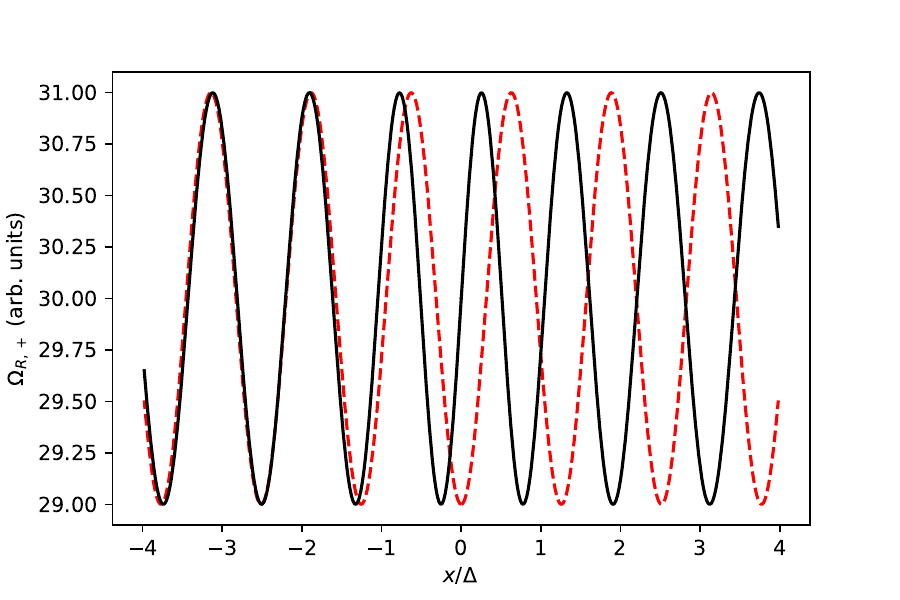}
    \caption{Plot of the Rabi frequency $\Omega_{R,+}$ in the magnetic domain wall system for $\Delta \, k=5$. The black line refers to the actual Rabi frequency $\Omega_{R,+}$ as given in Eq.~\eqref{Equation Rabi Frequencies Domain Wall}, whereas the dashed red line corresponds to $f(x) = | H^{\rm ref}_{x'} + H^{\rm ref}_{y'}  - \cos(k x) | $ and is plotted here to explicitly highlight the phase shift by $\pi$ of the Rabi frequency $\Omega_{R,+}$. Here we used $\theta_{N} = \phi_{N} = 0$, $H^{\rm ref}_{x'} = 20$ (arb. units) and $H^{\rm ref}_{y'}=10$ (arb. units). The parameter $|m|$ was chosen such that $M^{dw}_k (z) = -1$ (arb. units) at the fixed coordinate $z<0$. }
    \label{Figure Domain Wall Rabi Frequency k=5}
\end{figure}

As shown in Appendix \ref{Appendix Domain Wall}, the technique described in Sec. \ref{Section General Theory} can now be used to calculate the spin wave magnetic field in those regions where $k\gg |\frac{d \chi}{d x}|, | \frac{d \theta}{d x} |$. For $k \gg \Delta^{-1}$ this condition is satisfied everywhere, whereas for $k \lesssim \Delta^{-1}$ it can always be satisfied sufficiently far away from the domain wall. We find
\begin{align}\label{Equation Magnetizing Field Domain Wall}
    \mathbf{H}(\mathbf{r},t) &= M_{k}^{dw}(z)
    \begin{pmatrix}
    \cos [ k x - \omega t + \varphi_z(x)]\\
    0\\
    - \text{sgn}(z)  \sin [ k x - \omega t + \varphi_z(x)] 
    \end{pmatrix}
    ,
\end{align}
where
\begin{align}
    M_{k}^{dw}(z) &= -\frac{M_s |m| k}{2} \sqrt{1+\Delta^2 k^2} \int_{-d}^{d} d z' e^{-k|z-z'|}, \label{Equation Mdw(z)} \\
    \varphi_z(x) &= \chi(x) - \text{sgn}(z) \theta(x). \label{Equation Varphi DW}
\end{align}
Using the same reference field as before, we then find that the corresponding Rabi frequencies are given by
\begin{align}\label{Equation Rabi Frequencies Domain Wall}
    \Omega_{R,\pm}(\mathbf{r}) &= \frac{\eta \mu_0}{\sqrt{2}}\biggr| H^{\rm ref}_{x'} \pm H_{y'}^{\rm ref} + \mathcal{A}_1^{dw}(z) \cos[k x + \varphi_z(x)] \nonumber\\ &\pm \big(\mathcal{A}_2^{dw}(z) \mp \mathcal{A}_3^{dw}(z) \big)\sin[k x + \varphi_z(x)]\biggr|.
\end{align}
Here we have again invoked the high reference field limit and we have introduced the following functions:
\begin{align}
    \mathcal{A}_{1}^{dw}(z) &= M_{k}^{dw}(z) \cos \theta_N \cos \phi_N , \nonumber\\
     \mathcal{A}_{2}^{dw}(z) &= -M_{k}^{dw}(z) \sin \phi_N, \nonumber \\
     \mathcal{A}_{3}^{dw}(z) &= -\text{sgn}(z) M_{k}^{dw}(z) \sin \theta_N.
\end{align}

Let us now briefly comment on our results. We have derived the solution, Eq.~\eqref{Equation Final Spin Wave Magnetization Domain Wall}, for a spin wave traveling straight through a single magnetic domain wall. We have shown that this spin wave acquires two phase contribution, namely $\theta(x)$, which is directly related to the domain wall geometry, and $\chi(x)$, which is given in Eq.~\eqref{Equation Dynamic Phase Domain Wall}. These phases can be determined separately by means of Rabi frequency measurements in the regions $z>0$ and $z<0$, as can be seen immediately from Eqs.~\eqref{Equation Varphi DW} and ~\eqref{Equation Rabi Frequencies Domain Wall}. 

In order for geometric phases to arise, the adiabatic limit needs to apply,  which here corresponds to $k \gg \Delta^{-1}$. From Fig. \ref{Figure Chi}, we immediately see that in this limit the phase $\chi(x)$ becomes a position-independent constant. In contrast, the angle $\theta(x)$, which characterizes the equilibrium magnetization direction, always changes by $\pi$ after crossing the domain wall. This phase shift of $\pi$ is directly observable in the spatial profile of the Rabi frequencies, as can be seen from Fig. \ref{Figure Domain Wall Rabi Frequency k=5}.

Although it is tempting to interpret this phase shift as constituting a geometric phase, one needs to be careful here. While it is true that this phase shift arises purely as a consequence of the system's geometry, we have not yet satisfied the closed loop requirement which would make it a true geometric phase. In the domain wall system the equilibrium magnetization $\mathbf{M}_0$ corresponds to the parameter vector $\mathbf{R}$. For a geometric phase to arise we thus require that $\mathbf{M}_0$ returns to its original value, which is possible only if there are \textit{two} domain walls present.

Let us assume then that we have two domain walls of respective widths $\Delta_1$ and $\Delta_2$, with the first domain wall rotating the equilibrium magnetization direction from $+\hat{\mathbf{x}}$ to $-\hat{\mathbf{x}}$ and the second domain wall rotating the equilibrium magnetization direction from $-\hat{\mathbf{x}}$ back to $+\hat{\mathbf{x}}$. Considering the adiabatic limit $k \gg \Delta^{-1}$, where we have defined $\Delta = \min(\Delta_1,\Delta_2)$, the geometric phase acquired by the spin wave will be equal to the solid angle subtended by the closed loop path followed by the equilibrium magnetization $\mathbf{M}_0$ \cite{Dugaev2005}. In the case of two domain walls that rotate $\mathbf{M}_0$ in the $xz$ plane, the geometric phase will thus be equal to zero for instance. However, when one of these domain walls is replaced by one which rotates $\mathbf{M}_0$ in the $xy$ plane, we would find a geometric phase of $\pi$ instead. Since NV center magnetometry can be used effectively to extract phase shifts induced by a single domain wall, it is obvious then that the geometric phase acquired in a system with two domain walls can be detected accurately by exactly the same methodology.

\subsection{Position-dependent anisotropy axes} \label{Subsection Precession of the Anisotropy Ellipse}
We now consider a planar magnetic thin film with position-dependent anisotropy axes. To be specific, we again take the film to be oriented parallel to the $xy$ plane and extending from $z=-d$ to $z=d$, with $d$ small. An external magnetic field $\mathbf{B}^{ext} = B \hat{\mathbf{z}}$ is present, which causes the equilibrium magnetization of the film to point in the $\hat{\mathbf{z}}$-direction. We also assume there is an in-plane magnetic anisotropy in the film, with the corresponding anisotropy coefficients $K_1$ and $K_2$ satisfying $K_1,K_2 > 0$ and $K_1 \neq K_2$. The coefficients $K_1$ and $K_2$ are associated with the axes $\mathbf{e}_1 = (\cos \psi,-\sin \psi,0)$ and $\mathbf{e}_2 = (\sin \psi, \cos \psi, 0)$ respectively. 

It has been shown \cite{Ruckriegel2020} that spin waves can acquire a geometric phase when the angle $\psi$ is allowed to vary as a function of position, or equivalently, when the anisotropy axes $\mathbf{e}_1$ and $\mathbf{e}_2$ become position-dependent. This result can be derived using a similar approach as applied in Sec. \ref{Subsection Magnetic Domain Walls}. One starts again with an energy functional, which in this case is given by
\begin{align}\label{Equation Energy Functional Precessing Ellipse}
    E[\mathbf{m}] &= \int d \mathbf{r}' \biggr[ - g \mathbf{m} \cdot \mathbf{B}^{ext} -\frac{J}{2} \mathbf{m} \cdot \nabla^2 \mathbf{m} \nonumber\\ &+ \frac{K_1}{2} \biggr( m_x \cos \psi(x) - m_y \sin \psi(x) \biggr)^2 \nonumber \\ &+ \frac{K_2}{2} \biggr( m_x \sin \psi(x) + m_y \cos \psi(x) \biggr)^2  \biggr].
\end{align}
Here $g>0$ is a coefficient describing the coupling between the magnetization and the external field, and the angle $\psi$ is assumed to depend only on $x$ for algebraic convenience. Solving the Landau-Lifshitz equation and applying the adiabatic limit, $|k|\gg\big|\frac{d \psi}{d x}\big|$, the magnetization of a spin wave with frequency $\omega$ and wave vector $\mathbf{k} = k \hat{\mathbf{x}}$ will then be given by \cite{Ruckriegel2020}
\begin{equation}\label{Equation Magnetization Precessing Anisotropy Ellipse}
    \mathbf{M}_{sw}(\mathbf{r},t)= \mathbf{M}_1(\mathbf{r},t) + \mathbf{M}_2(\mathbf{r},t),
\end{equation}
where
\begin{align}
    \frac{\mathbf{M}_1(\mathbf{r},t)}{M(z)} &= \sqrt[\leftroot{-2}\uproot{2} 4]{\frac{\omega_2(k)}{\omega_1(k)}}
    \begin{pmatrix}
         \hspace{0.25cm}\cos \psi(x) \cos[k x - \omega t + \varphi(x)]  \\
          -  \sin \psi(x) \cos[k x - \omega t + \varphi(x)] \\
         0
    \end{pmatrix}
    ,
\end{align}
and
\begin{align}
    \frac{\mathbf{M}_2(\mathbf{r},t)}{-M(z)} &=  \sqrt[\leftroot{-2}\uproot{2} 4]{\frac{\omega_1(k)}{\omega_2(k)}}
    \begin{pmatrix}
         \sin\psi(x) \sin[k x - \omega t + \varphi(x)]  \\
          \cos \psi(x) \sin[k x - \omega t + \varphi(x)] \\
         0
    \end{pmatrix}.
\end{align}
The dispersion relation is given by $\omega(k) = \sqrt{\omega_1(k) \omega_2(k)}$, with
\begin{align}
     \omega_1(k) &= \gamma (J k^2 + g M_s B^{ext} + K_1)/M_s,  \\
     \omega_2(k) &= \gamma (J k^2 + g M_s B^{ext}  + K_2)/M_s.
\end{align}
The phase $\varphi(x)$ can be expressed in terms of $\psi(x)$ in the following way:
\begin{equation}
    \varphi(x) =   \frac{-2}{\sqrt{\frac{\omega_1(k)}{\omega_2(k)}} + \sqrt{\frac{\omega_2(k)}{\omega_1(k)}}} \int_0^x \psi'(x') d x'.
\end{equation}
Finally, assuming exchange boundary conditions, the function $M(z)$ is given by
\begin{equation}
    M(z) = M_s m \biggr[ H(z+d)-H(z-d)\biggr],
\end{equation}
with $m\in\mathbb{R}$ an arbitrary constant satisfying $|m|\ll 1$.

\begin{figure}
    \centering
    \includegraphics[scale=0.75]{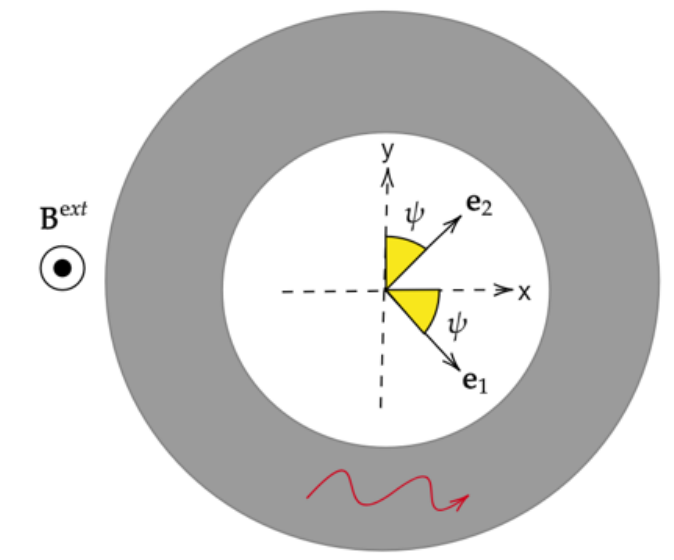}
    \caption{A ferromagnetic ring, confined to the $xy$-plane, in the presence of an external magnetic field $\mathbf{B}^{\mathrm{ext}}$ that points in the $\hat{\mathbf{z}}$-direction. The ring is assumed to have different anisotropy coefficients $K_1$ and $K_2$ along the (position-dependent) anisotropy axes $\mathbf{e}_1$ and $\mathbf{e}_2$. Spin waves traveling along this ring acquire a geometric phase because the directions of $\mathbf{e}_1$ and $\mathbf{e}_2$ are simultaneously changing.}
    \label{Figure Magnetic Ring}
\end{figure}

Now, similar to the domain wall system, the phase $\varphi(x)$ becomes a geometric phase whenever $\psi(x)$ returns to its original value modulo $2 \pi$. As suggested in Ref. \cite{Ruckriegel2020}, the system discussed in this section can be realized experimentally in ring structures made of magnetic materials with a large in-plane anisotropy. A sketch of such a system is shown in Figure \ref{Figure Magnetic Ring}.

To conclude this section, we will demonstrate that $\varphi(x)$ can also be measured directly by means of NV center magnetometry. As shown in Appendix \ref{Appendix Precessing Anisotropy}, the spin wave magnetic field and the corresponding NV center Rabi frequencies can again be calculated using the technique described in Section \ref{Section General Theory}. To avoid complexities of a purely algebraic nature we will not derive a completely general solution for the Rabi frequencies, but instead give the expressions for two limiting cases.

\textbf{Case (i.)}: $\omega_1 (k) \ll \omega_2(k)$. We find that the Rabi frequencies are given by
\begin{align}
    \Omega_{R,\pm}(\mathbf{r}) &= \frac{\eta \mu_0}{\sqrt{2}}\biggr| H^{\rm ref}_{x'} \pm H_{y'}^{\rm ref} + \mathcal{A}_1^{pa}(x,z) \cos[k x + \varphi(x)] \nonumber\\ &\pm \big(\mathcal{A}_2^{pa}(x,z) \mp \mathcal{A}_3^{pa}(x,z) \big)\sin[k x + \varphi(x)]\biggr|,
\end{align}
where
\begin{align}
    \mathcal{A}_{1}^{pa}(x,z) &= \text{sgn}(k) H_{12}(z) \cos \psi(x) \cos \theta_N \cos \phi_N , \nonumber\\
     \mathcal{A}_{2}^{pa}(x,z) &= -\text{sgn}(k) H_{12}(z) \cos \psi(x) \sin \phi_N, \nonumber \\
     \mathcal{A}_{3}^{pa}(x,z) &= -\text{sgn}(z) H_{12}(z) \cos \psi(x) \sin \theta_N.
\end{align}
Here we have introduced the function $H_{12}(z)$, which takes the following form:
\begin{align}
    H_{12}(z) = - \frac{M_s m k}{2} \sqrt{\frac{\omega_2(k)}{\omega_1(k)}} \int^d_{-d} d z' e^{-|k||z-z'|}.
\end{align}

\textbf{Case (ii.)}: $\omega_1 (k) \gg \omega_2(k)$. We find that the Rabi frequencies are given by
\begin{align}
    \Omega_{R,\pm}(\mathbf{r}) &= \frac{\eta \mu_0}{\sqrt{2}}\biggr| H^{\rm ref}_{x'} \pm H_{y'}^{\rm ref} + \mathcal{B}_1^{pa}(x,z) \cos[k x + \varphi(x)] \nonumber\\ &\pm \big(\mathcal{B}_2^{pa}(x,z) \mp \mathcal{B}_3^{pa}(x,z) \big)\sin[k x + \varphi(x)]\biggr|,
\end{align}
where
\begin{align}
    \mathcal{B}_{1}^{pa}(x,z) &= \text{sgn}(k) H_{21}(z) \sin \psi(x) \cos \theta_N \cos \phi_N , \nonumber\\
     \mathcal{B}_{2}^{pa}(x,z) &= -\text{sgn}(k) H_{21}(z) \sin \psi(x) \sin \phi_N, \nonumber \\
     \mathcal{B}_{3}^{pa}(x,z) &= -\text{sgn}(z) H_{21}(z) \sin \psi(x) \sin \theta_N.
\end{align}
Here we have introduced the function $H_{21}(z)$, which takes the following form:
\begin{align}
    H_{21}(z) = - \frac{M_s m k}{2} \sqrt{\frac{\omega_1(k)}{\omega_2(k)}} \int^d_{-d} d z' e^{-|k||z-z'|}.
\end{align}

From these expressions, we conclude that the geometric phase can be straightforwardly measured.

\section{Conclusion}\label{Section Conclusion}
In this paper we have demonstrated the promising potential of NV center magnetometry for the study of geometric phases in spin waves. Although optical techniques such as phase-resolved Brillouin light scattering \cite{Serga2006} have been used to measure spin wave phases \cite{Han2019,Wojewoda2020}, their spatial resolution is diffraction-limited to length scales on the order of $100$ nm. Contrastingly, NV center magnetometry can in principle offer a spatial resolution on the nanoscale \footnote{We note that the spatial resolution required to observe the spin wave geometric phase in a magnetometry measurement needs to be smaller than the wavelength of the spin wave.}, as here the limitations imposed on the resolution depend only on the distance between the NV center and the sample under investigation \cite{Taylor2008}. Harnessing the magnetic field sensing capabilities of single NV centers would thus enable the study of spin waves with nanometer wavelengths via a technique that, at least from a theoretical point of view, is straightforward to interpret.

Despite the fact that our calculations started from a relatively simple two-component form for the spin wave magnetization, Eq.~\eqref{Equation magnetization}, we note that more complicated expressions can easily be accommodated by a similar methodology. Although we assumed in Section \ref{Section General Theory} that our magnetic film was shaped in the form of an infinite plane, this assumption can also be relaxed significantly, since we only require that it appears like an infinite plane from the point of view of the NV center. By placing the NV center sufficiently close to the film, this condition can almost always be satisfied. Finally, we mention that we have ignored the effects on the phase extraction procedure that result from spin-wave damping. However, given that damping typically only gives rise to an exponential decay in the spin wave magnitude as a function of position, we expect that for sufficiently small spin wave damping, such as can be achieved in YIG-films \cite{Serga2010}, we can reliably extract the phase with the NV-center based detection technique described in this paper. 

Finally, the technique we have proposed to directly measure geometric phases for spin waves is not restricted to the two examples we discussed here. In principle, it could be used to directly detect the phase of the Bloch functions of a spin wave in a periodic structure - e.g. a magnonic crystal \cite{Gulyaev2003} -  thereby providing direct access to the Berry curvature and topology of spin wave band structure.

\begin{acknowledgments}
This work was supported by the Dutch Research Council (NWO) by the research programme Fluid Spintronics with Project No. 182.069 and by OCENW.XL21.XL21.058.

\end{acknowledgments}

\appendix 

\section{General Rabi frequency derivations}\label{Appendix Rabi Frequency}

\subsection{Derivation of Eq. \eqref{Equation Rabi Frequency Formula}}
We reproduce the derivation of Eq.~\eqref{Equation Rabi Frequency Formula}, starting from the ground state Hamiltonian,
\begin{equation}
    \hat{\mathcal{H}} = D \hat{S}_{z'}^2 + \eta \mathbf{B} \cdot \mathbf{S}.
\end{equation} 
The magnetic field is given by:
\begin{equation}
    \mathbf{B} = (B_R + B_L) \cos(\omega t) \hat{\mathbf{x}}' + (B_R - B_L) \sin(\omega t) \hat{\mathbf{y}}' + B_0 \hat{\mathbf{z}}',
\end{equation}
with $|B_0| \gg \sqrt{B_R^2+B_L^2}$. The $3\times 3$ Hamiltonian $\hat{\mathcal{H}}$ can be expressed in the basis $\{\ket{m_{z'}}\}$ ($m_{z'} = -1,0,1$), yielding
\begin{equation}\label{Equation Matrix Representation Hamiltonian}
    \hat{\mathcal{H}} =
    \begin{pmatrix}
    D+\eta B_0 & \frac{\eta}{\sqrt{2}}(B_{x'}-iB_{y'}) & 0 \\
    \frac{\eta}{\sqrt{2}}(B_{x'}+iB_{y'}) & 0 & \frac{\eta}{\sqrt{2}}(B_{x'}-iB_{y'}) \\
    0 & \frac{\eta}{\sqrt{2}}(B_{x'}+iB_{y'}) & D-\eta B_0
    \end{pmatrix}
\end{equation}
where $B_{x'} = B_R+B_L$ and $B_{y'} = B_R-B_L$. Although we are formally dealing with a single three-level Hamiltonian, we are allowed to split it into two effective two-level Hamiltonians $\hat{\mathcal{H}_{\pm}}$, with $\mathcal{H}_+$ ($\hat{\mathcal{H}_-}$) consisting of the levels $\ket{0}$ and $\ket{1}$ ($\ket{-1}$). This is a consequence of the fact that the transitions $\ket{1} \leftrightarrow \ket{-1}$ are dipole-forbidden, as discussed in Section \ref{Section NV Center Magnetometry}. We then have \cite{Bertelli2021}
\begin{align}
    \hat{\mathcal{H}}_{\pm} = &\frac{\omega_{\pm}}{2} \left(\hat{I}\pm\hat{\sigma}_{z'} \right) + \frac{\eta}{\sqrt{2}} B_{x'} \cos(\omega t) \hat{\sigma}_{x'} \nonumber\\&+ \frac{\eta}{\sqrt{2}} B_{y'} \sin(\omega t) \hat{\sigma}_{y'}, 
\end{align}
with $\omega_{\pm} = D \pm \eta B_0$ and $\{\sigma_{x'},\sigma_{y'},\sigma_{z'} \}$ the Pauli spin-$1/2$ matrices. We note that the Hamiltonians $\hat{\mathcal{H}}_{\pm}$ are periodic in time with period $T = \frac{2 \pi}{\omega}$.

To obtain the Rabi frequencies $\Omega_{R,\pm}$, one uses the standard Floquet theory approach \cite{Santoro}. Floquet's theorem states that the solution to the time-dependent Schr\"odinger equation,
\begin{equation}
    i \frac{d}{d t} \ket{\Psi(t)} = \hat{\mathcal{H}}(t) \ket{\Psi(t)},
\end{equation}
with a time-periodic Hamiltonian, $\hat{\mathcal{H}}(t+T) = \hat{\mathcal{H}}(t)$, can be written as
\begin{equation}
    \ket{\Psi(t)} = \hat{U}(t) \ket{\Psi(0)},
\end{equation}
where $\hat{U}(t)$ is given by:
\begin{equation}
    \hat{U}(t) = \hat{P}(t)e^{-i \hat{\mathcal{H}}_F t}.
\end{equation}
In the above expression we have introduced the Floquet Hamiltonian $\hat{\mathcal{H}}_F$, which is a time-independent Hermitian operator, and the stroboscopic kick operator $\hat{P}(t)$, which is a unitary operator satisfying
\begin{equation}\label{Equation Stroboscopic Kick Operator}
   \hat{P}(t+T) = \hat{P}(t).
\end{equation}
The Floquet Hamiltonian can be expressed in terms of $\hat{\mathcal{H}}(t)$ and $\hat{P}(t)$ as follows \cite{Santoro}:
\begin{equation}\label{Equation Floquet Hamiltonian}
    \hat{\mathcal{H}}_{F} = \hat{P}^{\dagger}(t) \hat{\mathcal{H}} (t) \hat{P}(t) - i  \hat{P}^{\dagger}(t) \frac{d \hat{P}(t)}{d t}.
\end{equation}
Solving for the (quasi)energies of the Floquet Hamiltonian allows us to determine the Rabi frequencies. 

We now apply the Floquet approach to our problem at hand, with $\hat{\mathcal{H}}(t) = \hat{\mathcal{H}}_{\pm}$. We seek to calculate the corresponding Floquet Hamiltonians $\hat{\mathcal{H}}_{F,\pm}$ using Eq.~\eqref{Equation Floquet Hamiltonian}, which requires us to know the stroboscopic kick operators $\hat{P}_{\pm}(t)$. Because there does not exist a general approach to find the stroboscopic kick operator, one has to make an educated guess. The first step is to guess a unitary
operator satisfying Eq.~\eqref{Equation Stroboscopic Kick Operator}. We use the following ansatz:
\begin{equation}\label{Equation Ansatz Kick Operator}
    \hat{P}_{\pm}(t) = e^{i \frac{\omega t}{2} (\hat{1}\mp\hat{\sigma}_{z'})}.
\end{equation}
If this makes the RHS of Eq.~\eqref{Equation Stroboscopic Kick Operator} time-independent, then the ansatz corresponds to the true stroboscopic kick operator. Plugging Eq.~\eqref{Equation Ansatz Kick Operator} into Eq.~\eqref{Equation Stroboscopic Kick Operator}, we get
\begin{align}
    \hat{\mathcal{H}}_{F,+} = &\frac{\omega + \omega_+}{2} \hat{I} - \frac{ \Delta_+}{2} \hat{\sigma}_z + \frac{\eta}{\sqrt{2}} B_R \hat{\sigma}_x \nonumber \\&+ \frac{\eta}{\sqrt{2}} B_L \biggr( \cos(2 \omega t) \hat{\sigma}_x -\sin (2 \omega t) \hat{\sigma}_y \biggr), \nonumber \\
     \hat{\mathcal{H}}_{F,-} = &\frac{\omega+\omega_-}{2} \hat{I} + \frac{ \Delta_-}{2} \hat{\sigma}_z + \frac{\eta}{\sqrt{2}} B_L \hat{\sigma}_x \nonumber \\&+ \frac{\eta}{\sqrt{2}} B_R \biggr( \cos(2 \omega t) \hat{\sigma}_x+\sin (2 \omega t) \hat{\sigma}_y \biggr),
\end{align}
where $\Delta_{\pm} = \omega - \omega_{\pm}$. It is obvious that the ansatz was not entirely correct, as there is still a time-dependency contained within the resulting expression for the Floquet Hamiltonians $\hat{\mathcal{H}}_{F,\pm}$. However, since the time-dependent terms oscillate rapidly with a frequency $2 \omega$, we may discard them by means of the rotating wave approximation \cite{Bertelli2021,Santoro}. We then end up with the following time-independent expressions for the Floquet Hamiltonians:
\begin{align}
    \hat{\mathcal{H}}_{F,+} = &\frac{\omega + \omega_+}{2} \hat{I} - \frac{ \Delta_+}{2} \hat{\sigma}_z + \frac{\eta}{\sqrt{2}} B_R \hat{\sigma}_x,   \nonumber\\
     \hat{\mathcal{H}}_{F,-} = &\frac{\omega+\omega_-}{2} \hat{I} + \frac{ \Delta_-}{2} \hat{\sigma}_z + \frac{\eta}{\sqrt{2}} B_L \hat{\sigma}_x.  
\end{align}
The (quasi)energies of $\hat{\mathcal{H}}_{F,\pm}$ are then:
\begin{align}
    \epsilon_{+} &= \frac{\omega + \omega_+}{2} \pm \frac{\Omega_+}{2}, \nonumber \\
    \epsilon_{-} &= \frac{\omega + \omega_-}{2} \pm \frac{\Omega_-}{2},
\end{align}
where the Rabi frequencies $\Omega_{\pm}$ are given by
\begin{align}
    \Omega_+ &= \sqrt{2 \eta^2 B_R^2 + \Delta_+^2}, \nonumber \\
    \Omega_- &= \sqrt{2 \eta^2 B_L^2 + \Delta_-^2}.
\end{align}
Letting either $\omega \rightarrow \omega_+$ or $\omega \rightarrow \omega_-$ we finally arrive at Eq.~\eqref{Equation Rabi Frequency Formula}.

\subsection{Derivation of Eq.~\eqref{Equation General Rabi Frequencies}}
We now derive the general expression for the NV Rabi frequencies, Eq.~\eqref{Equation General Rabi Frequencies}, resulting from the combined field $\mathbf{H}_I(\mathbf{r},t) = \mathbf{H} (\mathbf{r},t) + \mathbf{H}_{\rm ref} (\mathbf{r},t)$, where the spin wave field $\mathbf{H} (\mathbf{r},t)$ and the reference field $\mathbf{H}_{\rm ref} (\mathbf{r},t)$ are given by respectively Eq.~\eqref{Equation Approximate Magnetizing Field} and Eq.~\eqref{Equation Reference Field}.

We first treat the simple case where $\theta_N = \phi_N = 0$. Applying the compound-angle trigonometric identities $\cos(a-b) = \cos(a) \cos(b) + \sin(a) \sin(b)$ and $\sin(a-b) = \cos(b) \sin(a) -\cos(a) \sin(b)$ we find that the $x$- and $y$-components of the interference field are given by:
\begin{align}
    H_{I,x}(\mathbf{r},t) = &\biggr[\mathcal{A}_1(\mathbf{r}) \cos[\mathbf{k} \cdot \boldsymbol{\rho}+\varphi(\boldsymbol{\rho})] + H^{\rm ref}_x\biggr] \cos(\omega t) \nonumber \\&+\mathcal{A}_1(\mathbf{r}) \sin[\mathbf{k} \cdot \boldsymbol{\rho}+\varphi(\boldsymbol{\rho})] \sin (\omega t), \nonumber \\
    H_{I,y}(\mathbf{r},t) = &\biggr[\mathcal{A}_2(\mathbf{r}) \sin[\mathbf{k} \cdot \boldsymbol{\rho}+\varphi(\boldsymbol{\rho})] + H^{\rm ref}_y\biggr] \sin (\omega t) \nonumber \\&+\mathcal{A}_2(\mathbf{r}) \cos[\mathbf{k} \cdot \boldsymbol{\rho}+\varphi(\boldsymbol{\rho})] \cos(\omega t).
\end{align}
Denoting the time average of $H^2_{I,\beta}$ over a full period $T = \frac{2 \pi}{\omega}$ by $\langle H^2_{I,\beta} \rangle $ and using the trigonometric identity $a_1 \sin(\omega t) + a_2 \cos(\omega t) = A \sin(\omega t + \xi)$, with $A = \sqrt{a_1^2 + a_2^2}$ and $\tan (\xi) = \frac{a_2}{a_1}$, we can further simplify the above expressions into the following form:
\begin{align}\label{Equation Interference Field Components}
    H_{I,x}(\mathbf{r},t) &= \sqrt{2 \langle H_{I,x}^2(\mathbf{r}) \rangle} \sin[\omega t + \xi_x(\mathbf{r})], \nonumber\\
    H_{I,y}(\mathbf{r},t) &= \sqrt{2 \langle H_{I,y}^2(\mathbf{r}) \rangle} \sin[\omega t + \xi_y(\mathbf{r})], 
\end{align}
where we have
\begin{align}
    \tan\big(\xi_x(\mathbf{r})\big) &= \frac{H^{\rm ref}_x+\mathcal{A}_1(\mathbf{r}) \cos[\mathbf{k}\cdot\boldsymbol{\rho}+\varphi(\boldsymbol{\rho})]}{\mathcal{A}_1(\mathbf{r}) \sin[\mathbf{k}\cdot\boldsymbol{\rho}+\varphi(\boldsymbol{\rho})]}, \nonumber \\ 
    \tan\big(\xi_y(\mathbf{r})\big) &= \frac{\mathcal{A}_2 (\mathbf{r}) \cos[\mathbf{k}\cdot\boldsymbol{\rho}+\varphi(\boldsymbol{\rho})]}{H^{\rm ref}_y+\mathcal{A}_2(\mathbf{r}) \sin[\mathbf{k}\cdot\boldsymbol{\rho}+\varphi(\boldsymbol{\rho})]}.
\end{align}
To obtain the Rabi frequencies we need to decompose $H_{I,x}$ and $H_{I,y}$ into right- and left-handed components. Due to the complicated positional dependency of the angles $\xi_x(\mathbf{r})$ and $\xi_y(\mathbf{r})$, which are directly related to the polarization of the interference field, this is usually rather difficult. Furthermore, an efficient extraction of $\varphi(\boldsymbol{\rho})$ from the Rabi frequency measurements will become increasingly challenging for rapidly varying $\xi_x(\mathbf{r})$ and $\xi_y(\mathbf{r})$. For these reasons we seek to eliminate the positional dependency of $\xi_x(\mathbf{r})$ and $\xi_y(\mathbf{r})$ altogether, which can be accomplished by going to the limit of large reference fields, $|H^{\rm ref}_{x}|\gg|\mathcal{A}_1(\mathbf{r})|$ and $|H^{\rm ref}_{y}|\gg|\mathcal{A}_2(\mathbf{r})|$. From $\tan(0) = \tan(\pi) = 0$ and 
\begin{align*}
    \lim_{\epsilon\to 0^+} \tan(-\pi/2+\epsilon) &= \lim_{\epsilon\to 0^+} \tan(\pi/2+\epsilon) = -\infty,\\
    \lim_{\epsilon\to 0^-} \tan(-\pi/2+\epsilon) &= \lim_{\epsilon\to 0^-} \tan(\pi/2+\epsilon) = \infty,
\end{align*}
we immediately find that for $H^{\rm ref}_x > 0$ ($H^{\rm ref}_x < 0$) we have $\xi_x \rightarrow \pi/2$ ($\xi_x \rightarrow -\pi/2$), whereas for $H^{\rm ref}_y > 0$ ($H^{\rm ref}_y < 0$) we have $\xi_y \rightarrow 0$ ($\xi_y \rightarrow \pi$) in this high reference field limit. With this significant simplification at hand it is now straightforward to decompose $H_{I,x}$ and $H_{I,y}$ into right- and left-handed components. Assuming $H^{\rm ref}_x$ and $H^{\rm ref}_y$ to have the same sign, we immediately find
\begin{align}
    \Omega_{R,+}(\mathbf{r}) &= \eta \mu_0 \biggr|   \sqrt{ \langle H_{I,x}^2(\mathbf{r}) \rangle}+\sqrt{ \langle H_{I,y}^2(\mathbf{r}) \rangle}\biggr|, \nonumber\\
    \Omega_{R,-}(\mathbf{r}) &= \eta \mu_0 \biggr|   \sqrt{ \langle H_{I,x}^2(\mathbf{r}) \rangle}-\sqrt{ \langle H_{I,y}^2(\mathbf{r}) \rangle}\biggr|.
\end{align}
These results can be simplified even further by applying the high reference field limit to $ \langle H_{I,x}^2(\mathbf{r}) \rangle^{1/2}$ and $ \langle H_{I,y}^2(\mathbf{r}) \rangle^{1/2}$. Using $\sqrt{1+a} = 1+\frac{a}{2}+\mathcal{O}(a^2)$ and dropping terms of order $\mathcal{O}(1/H^{\rm ref})$ we then obtain:
\begin{align} 
    \Omega_{R,\pm}(\mathbf{r}) &= \frac{\eta \mu_0}{\sqrt{2}}\biggr| H^{\rm ref}_{x'} \pm H_{y'}^{\rm ref} + \mathcal{A}_1(\mathbf{r}) \cos[\mathbf{k}\cdot \boldsymbol{\rho}+\varphi(\boldsymbol{\rho})] \nonumber\\ &\pm \mathcal{A}_2(\mathbf{r}) \sin[\mathbf{k}\cdot \boldsymbol{\rho}+\varphi(\boldsymbol{\rho})]\biggr|.
\end{align}
Exactly the same methodology can now be applied to the case where $\theta_N$ and $\phi_N$ take on arbitrary values, which will then immediately lead to Eq.~\eqref{Equation General Rabi Frequencies}. 

\subsection{First-order corrections due to small polarization angles $\delta \xi$}
In the previous section we showed that the position dependency of the angles $\xi_x(\mathbf{r})$ and $\xi_y(\mathbf{r})$ can be eliminated by going to the high reference field limit. Strictly speaking, this dependency only disappears in the limit where $|H^{\rm ref}_x|$ and $|H^{\rm ref}_y|$ tend to infinity, and we should therefore investigate the corrections that arise due to the reference field components having large but finite values. Experimental limitations aside, there are also theoretical reasons why infinitely large reference fields are problematic for our model. First of all, too large a reference field will make the spatial contrast in the interference field too small to detect the effects of the phase $\varphi (\boldsymbol{\rho})$. The second reason is more subtle and revolves around the critical assumption made in the rotating wave approximation that $\Omega_{R,+}$ ($\Omega_{R,-}$) is much less than the resonance frequency $\omega_+$ ($\omega_-$) \cite{Wu2007}. As the Rabi frequencies increase with the magnitude of the time-varying magnetic field, it is clear that for extremely large reference fields the model used in this paper will break down.

Let us now consider the first-order corrections to Eq.~\eqref{Equation General Rabi Frequencies} that arise due to the finite size of the reference field magnitude. For these purposes we first rewrite Eq.~\eqref{Equation Interference Field Components} into a slightly more convenient form. Changing the origin of time such that $\omega t \rightarrow \omega t - \xi_x (\mathbf{r}) + \pi/2$ and introducing the angle $\xi(\mathbf{r}) = \xi_y(\mathbf{r})-\xi_x(\mathbf{r})+\pi/2$, we get the following simplified expressions for Eq.~\eqref{Equation Interference Field Components}:
\begin{align}
    H_{I,x}(\mathbf{r},t) &= \sqrt{2 \langle H_{I,x}^2(\mathbf{r}) \rangle} \cos(\omega t), \nonumber\\
    H_{I,y}(\mathbf{r},t) &= \sqrt{2 \langle H_{I,y}^2(\mathbf{r}) \rangle} \sin[\omega t + \xi(\mathbf{r})].
\end{align}
We now focus on the case where $H^{\rm ref}_x$ and $H^{\rm ref}_y$ have the same sign such that $\xi(\mathbf{r} ) \rightarrow 0$ in the high reference field limit, and we calculate the first-order correction to Eq.~\eqref{Equation General Rabi Frequencies} in terms of small deviations $\delta \xi$ around $\xi = 0$. We note that the case where $H^{\rm ref}_x$ and $H^{\rm ref}_y$ have different signs will lead to the same conclusion regarding this first-order correction.

To calculate the correction we start from the following Hamiltonians: 
\begin{align}\label{Equation Hamiltonian with Polarization Angle}
    \hat{\mathcal{H}}_{\pm} = &\frac{\omega_{\pm}}{2} (\hat{I}\pm\hat{\sigma}_{z}) + \frac{\eta}{\sqrt{2}}  \cos(\omega t) (B_R+B_L) \hat{\sigma}_{x} \nonumber \\&+ \frac{\eta}{\sqrt{2}} \sin(\omega t + \delta \xi) (B_R-B_L) \hat{\sigma}_{y}.
\end{align}
Working to first order in $\delta \xi$, we have $\sin(\omega t + \delta \xi) = \sin(\omega t) + \delta\xi \cos(\omega t)$. Plugging this into Eq.~\eqref{Equation Hamiltonian with Polarization Angle} and applying the rotating wave approximation to discard all terms oscillating with frequency $2 \omega$, we find that the change $\Delta \hat{\mathcal{H}}_{F,\pm}$ of the corresponding Floquet Hamiltonians is given to first order in $\delta \xi$ by:
\begin{equation}
    \Delta \hat{\mathcal{H}}_{F,\pm} = \frac{\eta}{2\sqrt{2}} \delta \xi (B_R-B_L) \hat{\sigma}_y.
\end{equation}
To obtain the first-order correction to the on-resonance Rabi frequencies $\Omega_{R,\pm}$ we then have to consider the quasi-energies of the following $2 \times 2$ Floquet Hamiltonians:
\begin{align}
    \hat{\mathcal{H}}_{F,+} &= \omega_+  + \frac{\eta}{\sqrt{2}} B_R \hat{\sigma}_x +  \Delta \hat{\mathcal{H}}_{F,+} \nonumber, \\
    \hat{\mathcal{H}}_{F,-} &= \omega_-  + \frac{\eta}{\sqrt{2}} B_L \hat{\sigma}_x +  \Delta \hat{\mathcal{H}}_{F,-}.
\end{align}
It immediately follows that the corrections to the quasi-energies are of order $\mathcal{O}(\delta \xi)^2$, and hence the first-order corrections in $\delta \xi$ to the Rabi frequencies $\Omega_{R,\pm}$ vanish. 

From this result we can thus conclude the following. As long as $\delta \xi$ is small enough such that we are allowed to approximate $\sin(\omega t + \delta \xi)$ by $\sin(\omega t) + \delta \xi \cos(\omega t)$, it is safe to use Eq.~\eqref{Equation General Rabi Frequencies} for the Rabi frequencies. Translating this into a condition on the reference field, we expect a reference field with its components about ten times larger than that of the spin wave field to already be sufficient for Eq.~\eqref{Equation General Rabi Frequencies} to hold.

\section{Magnetic field due to a spin wave traveling through a domain wall}\label{Appendix Domain Wall}
We derive the expression for the spin wave magnetic field, Eq.~\eqref{Equation Magnetizing Field Domain Wall}, starting from
Eq.~\eqref{Equation Final Spin Wave Magnetization Domain Wall}. The NV center is assumed to be placed sufficiently close to the magnetic film, such that the film's shape can be treated as that of an infinite plane. We first go back to the general formula:
\begin{equation}
    H_{\mathbf{\beta}} (\mathbf{r},t) = \frac{1}{4 \pi} \partial_{\beta} \int d\mathbf{r}' \frac{\nabla ' \cdot \mathbf{M}(\mathbf{r}',t)}{|\mathbf{r}-\mathbf{r}'|}.
\end{equation}
Using the identities $\cos(a) \cos(b) = \frac{1}{2} [ \cos(a+b)+\cos(a-b)]$ and $\sin(a) \cos(b) = \frac{1}{2} [ \sin(a+b)+\sin(a-b)]$, it is obvious that the same magnetic field would be generated by a magnetization $\mathbf{M}_{alt}(\mathbf{r},t)$ given by
\begin{equation}
   \mathbf{M}_{alt}(\mathbf{r},t) = \mathbf{M}_{alt,1}(\mathbf{r},t) + \mathbf{M}_{alt,2}(\mathbf{r},t),
\end{equation}
where
\begin{align}
    \mathbf{M}_{alt,1}(\mathbf{r},t) &= \frac{M(x,z)}{2}
    \begin{pmatrix}
        \cos [ k x - \omega t + \varphi_1(x)] \\
        0 \\
        \sin  [ k x - \omega t + \varphi_1(x)]
    \end{pmatrix},
    \nonumber
    \\
    \mathbf{M}_{alt,2}(\mathbf{r},t) &= \frac{M(x,z)}{2}
    \begin{pmatrix}
        \cos [ k x - \omega t + \varphi_2(x)] \\
        0 \\
        -\sin  [ k x - \omega t + \varphi_2(x)]
    \end{pmatrix}.
\end{align}
Here we have introduced the angles $\varphi_1(x)$ and $\varphi_2(x)$, which are defined as follows:
\begin{align}
    \varphi_1(x) &= \chi(x) - \theta(x), \nonumber \\
    \varphi_2(x) &= \chi(x) + \theta(x).
\end{align}
The similarities with the setup of Section \ref{Section General Theory} should be obvious; the only significant difference being that we now have an $x$-dependency contained in the function $M(x,z)$, which disappears in those regions where $k \gg |\frac{d \chi}{d x}|, |\frac{d \theta}{d x}|$. For wavelengths that are much smaller than the domain wall width $\Delta$, this condition on the wave vector k is satisfied everywhere, whereas for wavelengths that are of a size larger than or similar to $\Delta$ it only holds in the region $|x| \gg \Delta$. Assuming we are in a region where $k \gg |\frac{d \chi}{d x}|, |\frac{d \theta}{d x}|$ and using the fact that the total spin wave magnetic field is given by $\mathbf{H} = \mathbf{H}_{alt,1} + \mathbf{H}_{alt,2}$, we then find:
\begin{align}
    \mathbf{H}(\mathbf{r},t) &= M_{k}^{dw}(z)
    \begin{pmatrix}
    \cos [ k x - \omega t + \varphi_z(x)]\\
    0\\
    - \text{sgn}(z)  \sin [ k x - \omega t + \varphi_z(x)] 
    \end{pmatrix},
\end{align}
where we have defined $M^{dw}_k(z)$ and $\varphi_z(x)$ in Eqs. (\ref{Equation Mdw(z)}) and (\ref{Equation Varphi DW}) respectively.

\section{Spin wave magnetic field of the system with position-dependent anisotropy axes}\label{Appendix Precessing Anisotropy}
We derive an expression for the spin wave magnetic field starting from Eq.~\eqref{Equation Magnetization Precessing Anisotropy Ellipse}, using the same method discussed in Appendix \ref{Appendix Domain Wall}. First, we observe that exactly the same magnetic field would be generated by a magnetization $\mathbf{M}_{alt}(\mathbf{r},t)$ given by
\begin{equation}
   \mathbf{M}_{alt}(\mathbf{r},t) = \mathbf{M}_{alt,1}(\mathbf{r},t) + \mathbf{M}_{alt,2}(\mathbf{r},t),
\end{equation}
where
\begin{align}
    \frac{\mathbf{M}_{alt,1}(\mathbf{r},t)}{  \biggr(\sqrt[\leftroot{-2}\uproot{2} 4]{\frac{\omega_2}{\omega_1}} - \sqrt[\leftroot{-2}\uproot{2} 4]{\frac{\omega_1}{\omega_2}} \biggr)} &= \frac{M(z)}{2} 
    \begin{pmatrix}
        \cos[k x - \omega t +\varphi_1(x)] \\
        0\\
        0
    \end{pmatrix},
\end{align}
and
\begin{align}
    \frac{\mathbf{M}_{alt,2}(\mathbf{r},t)}{\biggr( \sqrt[\leftroot{-2}\uproot{2} 4]{\frac{\omega_2}{\omega_1}} + \sqrt[\leftroot{-2}\uproot{2} 4]{\frac{\omega_1}{\omega_2}} \biggr)} &= \frac{M(z)}{2}
    \begin{pmatrix}
        \cos[k x - \omega t +\varphi_2(x)] \\
        0\\
        0
    \end{pmatrix}.
\end{align}
Here we have introduced the angles $\varphi_1(x)$ and $\varphi_2(x)$, which are defined as follows:
\begin{align}
    \varphi_1(x) &= \varphi(x) - \psi(x), \nonumber\\
    \varphi_2(x) &= \varphi(x) + \psi(x).
\end{align}
Because the spin wave magnetization in Eq.~\eqref{Equation Magnetization Precessing Anisotropy Ellipse} was already derived under the assumption that the wave vector satisfies $|k| \gg|\frac{d \psi}{d x}|$, we can immediately use the results of Section \ref{Section General Theory} to obtain the spin wave magnetic field $\mathbf{H}$. We find $\mathbf{H} = \mathbf{H}_{alt,1} + \mathbf{H}_{alt,2}$, with
\begin{align}
    \mathbf{H}_{alt,1}(\mathbf{r},t) &= M_{k,1}^{pa}(z)
    \begin{pmatrix}
        \text{sgn}(k) \cos[k x - \omega t +\varphi_1(x)] \\
        0\\
        -\text{sgn}(z) \sin[k x - \omega t + \varphi_1(x)]
    \end{pmatrix},
\end{align}
and
\begin{align}
     \mathbf{H}_{alt,2}(\mathbf{r},t) &= M_{k,2}^{pa}(\mathbf{r})
    \begin{pmatrix}
       \text{sgn}(k) \cos[k x - \omega t +\varphi_2(x)] \\
        0\\
        -\text{sgn}(z) \sin[k x - \omega t + \varphi_2(x)]
    \end{pmatrix}.
\end{align}
In these expressions we have introduced the following definitions:
\begin{align}
    M_{k,1}^{pa}(z) &= - \frac{M_s m k}{4}\biggr( \sqrt[\leftroot{-2}\uproot{2} 4]{\frac{\omega_2}{\omega_1}} - \sqrt[\leftroot{-2}\uproot{2} 4]{\frac{\omega_1}{\omega_2}} \biggr) \int_{-d}^d d z' e^{-|k||z-z'|}, \nonumber\\
     M_{k,2}^{pa}(z) &= - \frac{M_s m k}{4} \biggr( \sqrt[\leftroot{-2}\uproot{2} 4]{\frac{\omega_2}{\omega_1}} + \sqrt[\leftroot{-2}\uproot{2} 4]{\frac{\omega_1}{\omega_2}} \biggr) \int^d_{-d} d z' e^{-|k||z-z'|} .
\end{align}

In the case where $\omega_1(k) \ll \omega_2(k)$, the spin wave magnetic field $\mathbf{H}$ takes the following simplified form:
\begin{align}
    \frac{\mathbf{H}(\mathbf{r},t)}{H_{12}(z)} \approx \cos\psi(x)
    \begin{pmatrix}
       \text{sgn}(k) \cos[k x - \omega t +\varphi(x)] \\
        0\\
        -\text{sgn}(z) \sin[k x - \omega t + \varphi(x)]
    \end{pmatrix}
    ,
\end{align}
where
\begin{align}
    H_{12}(z) = - \frac{M_s m k}{2} \sqrt{\frac{\omega_2(k)}{\omega_1(k)}} \int^d_{-d} d z' e^{-|k||z-z'|}.
\end{align}
Similarly, for $\omega_2(k) \ll \omega_1(k)$, we find
\begin{align}
    \frac{\mathbf{H}(\mathbf{r},t)}{H_{21}(z)} \approx \sin\psi(x)
    \begin{pmatrix}
       -\text{sgn}(k) \sin[k x - \omega t +\varphi(x)] \\
        0\\
        -\text{sgn}(z) \cos[k x - \omega t + \varphi(x)]
    \end{pmatrix}
    ,
\end{align}
where
\begin{align}
    H_{21}(z) = - \frac{M_s m k}{2} \sqrt{\frac{\omega_1(k)}{\omega_2(k)}} \int^d_{-d} d z' e^{-|k||z-z'|}.
\end{align}
As a minor comment, we mention that a redefinition of the time origin allows us to write the result for the limit $\omega_2(k) \ll \omega_1(k)$ as:
\begin{align}
    \frac{\mathbf{H}(\mathbf{r},t)}{H_{21}(z)} \approx \sin\psi(x)
    \begin{pmatrix}
       \text{sgn}(k) \cos[k x - \omega t +\varphi(x)] \\
        0\\
        -\text{sgn}(z) \sin[k x - \omega t + \varphi(x)]
    \end{pmatrix}
    ,
\end{align}
The Rabi frequencies of these fields can then be determined straightforwardly by means of Eq.~\eqref{Equation General Rabi Frequencies}.


\bibliography{biblio}

\end{document}